\documentclass[conference]{IEEEtran}
\IEEEoverridecommandlockouts
\usepackage{cite}
\usepackage{amsmath,amssymb,amsfonts}
\usepackage{algorithmic}
\usepackage{graphicx}
\usepackage{textcomp}
\usepackage{xcolor}
\def\BibTeX{{\rm B\kern-.05em{\sc i\kern-.025em b}\kern-.08em
    T\kern-.1667em\lower.7ex\hbox{E}\kern-.125emX}}

\usepackage{microtype}
\usepackage{graphicx}
\usepackage{subfigure}
\usepackage{booktabs} 
\usepackage{multirow} 
\usepackage{float}
\usepackage{amssymb}
\usepackage{adjustbox}
\usepackage{caption}
\usepackage{array}
\usepackage{multicol}
\usepackage{amsmath}
\usepackage{lscape}
\usepackage{tcolorbox}
\usepackage{xspace}
\usepackage{enumitem}
\usepackage{pgfplots}
\usepackage{subcaption}
\usepackage{hyperref}
\usepackage{orcidlink} 
\usepackage{xcolor}      
\pgfplotsset{compat=1.17}
\usepackage[numbers,square,sort&compress]{natbib}

\newcommand{\etal}{\emph{et al.}\xspace}
\renewcommand{\baselinestretch}{0.97}

\begin{document}

\title{Fact-Checking with Contextual Narratives: Leveraging Retrieval-Augmented LLMs for Social Media Analysis}

\author{
Arka Ujjal Dey~\orcidlink{0000-0001-8392-1574}, 
Muhammad Junaid Awan~\orcidlink{0000-0002-3857-1293}, 
Georgia Channing~\orcidlink{0009-0001-6354-7527}, 
Christian Schroeder de Witt~\orcidlink{0000-0003-4245-1179}, \\
and John Collomosse~\orcidlink{0000-0003-3580-4685}
}

\maketitle

\begin{abstract}
We propose CRAVE (Cluster-based Retrieval Augmented Verification with Explanation); a novel framework that integrates retrieval-augmented Large Language Models (LLMs) with clustering techniques to address fact-checking challenges on social media. CRAVE automatically retrieves multimodal evidence from diverse, often contradictory, sources.  Evidence is clustered into coherent narratives, and evaluated via an LLM-based judge to deliver fact-checking verdicts explained by evidence summaries. By synthesizing evidence from both text and image modalities and incorporating agent-based refinement, CRAVE ensures consistency and diversity in evidence representation. Comprehensive experiments demonstrate CRAVE's efficacy in retrieval precision, clustering quality, and judgment accuracy, showcasing its potential as a robust decision-support tool for fact-checkers.
\end{abstract}
\maketitle
\section{Introduction}

Fact-checking has emerged as a critical task in the era of social media~\citep{shahid2022detecting}, where information spreads rapidly~\citep{vosoughi2018spread,aimeur2023fake} and misinformation can have far-reaching consequences~\citep{kshetri2017economics,caceres2022impact,allcott2017social}. Automated fact-checking systems have gained traction as scalable solutions, yet they often face challenges such as handling diverse evidence sources, integrating multimodal data, and presenting comprehensive narratives. Traditional approaches~\citep{papadopoulos2024reddotmultimodalfactcheckingrelevant,papadopoulos2024similarity} focus on verifying factual claims against a static repository of information~\citep{papadopoulos2024verite,luo2021newsclippings,abdelnabi2022open,tonglet2025cove}, which may overlook nuances in evidence and the broader social context in which a claim spreads~\citep{gong2024integrating}. 

Recent advances in Large Language Models (LLMs) and Retrieval-Augmented Generation (RAG) have opened new avenues for building robust fact-checking frameworks. RAG has been shown to be effective in reducing factual inconsistencies ("hallucinations") during text generation, making it attractive for the fact-checking task~\citep{lewis2020retrieval,momii2024rag}; this is a crucial advantage, as such errors have been shown to significantly erode user trust in AI systems~\citep{amaro2023believe}. RAG models can access dynamic external knowledge, reason over complex evidence, and produce context-aware summaries. However, effectively leveraging these capabilities requires addressing several challenges: (1) how to retrieve relevant evidence efficiently, (2) how to cluster and organize potentially contradictory bodies of evidence into coherent narratives, and (3) how to generate holistic evaluations that incorporate these multiple perspectives.

\begin{figure}[h]
    \centering
    \includegraphics[width=\columnwidth]{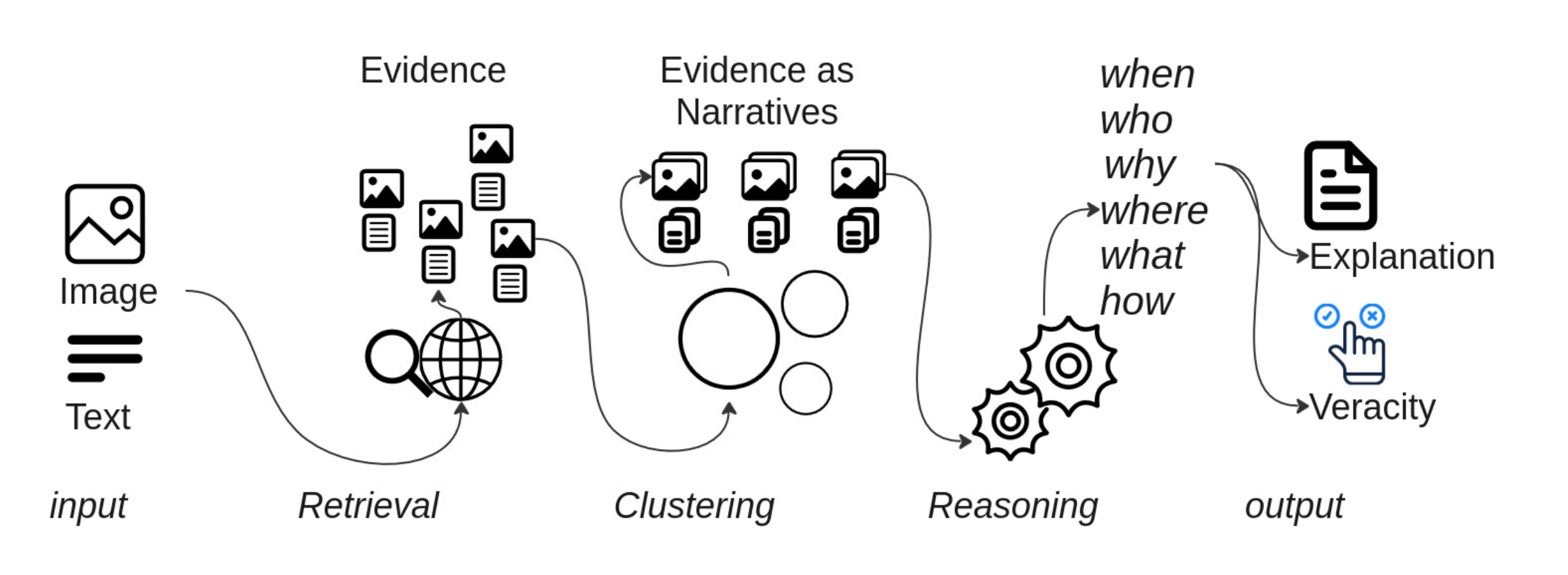}
    \caption{CRAVE is a fact-checking method that analyses multimodal claims, expressed in social media posts,  to determine their veracity.  It retrieves multimodal evidence, clusters it into distinct narratives, and uses an LLM with 5W1H  (who, what, when, where, why, and how) reasoning to produce an interpretable explanation and veracity verdict. }\label{fig:main}
\end{figure}

In this work, we propose CRAVE (Cluster-based Retrieval Augmented Verification with Explanation); a novel framework, \autoref{fig:main}, that integrates retrieval-augmented LLMs with clustering techniques to identify, structure, and evaluate evidence for fact-checking social media posts. The framework is designed to process multi-modal inputs (text and images) and iteratively refine evidence through agent-based mechanisms. By clustering evidence into distinct narratives and identifying external support (or otherwise) for those narratives, CRAVE enables the assessment of contentious topics that present with support from contradictory and noisy news sources.  The contributions of CRAVE are threefold:

{\bf 1. Narrative retrieval pipeline.} We introduce a modular pipeline integrating reverse image search, LLM-assisted text retrieval, and clustering to extract and organize evidence into narratives relevant to the claim to be checked.   \\
{\bf 2. Clustering and narrative extraction.} We propose a clustering strategy to group evidence into distinct narratives, followed by agent-based refinement. Each cluster’s narrative undergoes iterative verification ensuring intra-cluster consistency to produce coherent bodies of evidence. By grouping conflicting sources into separate clusters, CRAVE preserves nuances among multiple perspectives rather than forcing the LLM to resolve all contradictions in a single prompt.\\
{\bf 3. Cluster-based explainable judgment.} An LLM judge synthesizes the refined clusters into a final verdict, explaining the decision using the multiple perspectives from the clustered narratives to highlight how narrative threads support or contradict the claim.

We validate CRAVE on multiple real-world and synthetic datasets, showing that breaking up evidence into narrative clusters improves both precision and recall of fact-checking decisions. Comprehensive experiments reveal that our cluster-based pipeline excels in retrieval precision, clustering quality, and final veracity judgments. Human evaluators also rank the explanations generated by CRAVE as more coherent and comprehensive than those generated by baseline systems.

\section{Related work}
Automated fact-checking systems deployed in practice commonly work by matching newly surfaced claims against a library of already fact-checked claims~\citep{graves2018understanding}.  While efficient for well-known misinformation that periodically resurfaces, such systems struggle with novel or emerging misinformation. Research to date has mostly focused upon detecting intrinsic artifacts within a claim~\citep{liu2024mmfakebench, shao2023detecting} using synthetically generated datasets~\citep{aneja2021cosmos,shao2023detecting,liu2024mmfakebench}.
Such black-box detection methods often lack the kind of social and contextual explainability~\citep{shahid2022detecting} needed for users to trust their verdicts~\citep{gong2024integrating}, while also lagging behind developments in generative AI,  limiting their effectiveness and on novel manipulation schemes.

Thus, while the detection of intrinsic artifacts within manipulated media—such as image edits, swaps~\citep{shao2023detecting}, or deepfakes~\citep{liu2024mmfakebench}—constitutes a valuable and active line of inquiry, we advocate for a paradigm shift towards external evidence-based verification. Rather than focusing on increasingly brittle and often ephemeral manipulation signatures within a given image, our approach prioritizes a comparative analysis against a corpus of external, verifiable sources. By leveraging robust image similarity measures, we can retrieve visually analogous content and, critically, analyze the surrounding context. This methodology allows for a more nuanced judgment: an image can be authenticated as ``used in context," flagged as ``out of context" (a prevalent form of misinformation), or identified as a likely fabrication (e.g., a deepfake) due to its conspicuous absence from trusted external sources or its prior identification and debunking within them. This external, context-aware framework offers a more resilient and interpretable alternative to the current adversarial dynamic of detection and evasion.

Evidence-driven methods and datasets have been introduced to enhance the explainability of fact-checking frameworks. However, the evidence used is often curated and sanitized~\citep{papadopoulos2024verite,abdelnabi2022open,sharma2024amir}. As a result, methods such as~\citep{papadopoulos2024similarity,papadopoulos2024reddotmultimodalfactcheckingrelevant} achieve state-of-the-art results while bypassing the challenge of retrieving relevant evidence, which limits their applicability to real-world scenarios~\citep{shahid2022detecting}. Although these approaches are evidence-based, they are primarily data-driven trained classifiers that struggle to generalize across novel claims. 

\subsection{Retrieval of evidence}
Effective verification of multimodal misinformation depends on locating reliable evidence, particularly for images reused out of context~\citep{abdelnabi2022open,papadopoulos2024reddotmultimodalfactcheckingrelevant,braun2025defamedynamicevidencebasedfactchecking}.  As identified in~\citep{graves2018understanding} evidence retrieval is still done primarily by human fact-checkers through manually constructing search terms.  Prior works have demonstrated that incorporating reverse image search to find an image’s provenance and text-based search for contextual information can similarly improve automated fact-checking performance~\citep{abdelnabi2022open, tonglet2025cove, dey2024retrieval, qi2024sniffer}. Abdelnabi \etal were among the first to show that combining reverse image search with text retrieval improves the scrutiny of multimedia claims. Since then, other pipelines have extended such concepts to incorporate more robust retrieval modules or specialized knowledge. In~\citep{tonglet2025cove} the authors extend their sources through WIKIDATA~\citep{vrandevcic2014wikidata} and Wikichat~\citep{semnani2023wikichat} to enrich their image context, but the authors also note the limited effectiveness in non-western contexts. 

In real-world scenarios, fact-checking is complicated by how social media users often combine verbose or biased commentary with images unrelated to mainstream coverage. To address this, Dey \etal~\citep{dey2024retrieval} introduced a feedback-driven approach in which an autonomous agent refines search terms iteratively to mimic a journalist’s evolving research strategy. Such iterative refinements allow the system to capture elusive evidence more effectively than a single-pass search, particularly when no direct textual match exists for the claim.

\subsection{Reasoning with evidence} 
Data-driven fact-checking struggles with generalization, even when using external evidence. Prior supervised fact-checking models often overfit to specific patterns in training data and underperform on new domains.  This was demonstrated by Tonglet \etal  \citep{tonglet2025cove} who showed that  state-of-the-art models \citep{papadopoulos2024similarity,papadopoulos2024reddotmultimodalfactcheckingrelevant} degrade significantly on real-world datasets compared to curated benchmarks, likely due to domain mismatch. Moreover, even if external evidence is retrieved, many approaches simply score claim-evidence alignment in a single pass based on ranking, ignoring contradictory or ambiguous data. In contrast to the use of heuristics to rank evidence~\citep{papadopoulos2024reddotmultimodalfactcheckingrelevant}, we propose clustering distinct narratives and applying judgment that accounts for support or conflict from each cluster. 

\begin{figure*}[h]
    \centering
    \includegraphics[width=\textwidth]{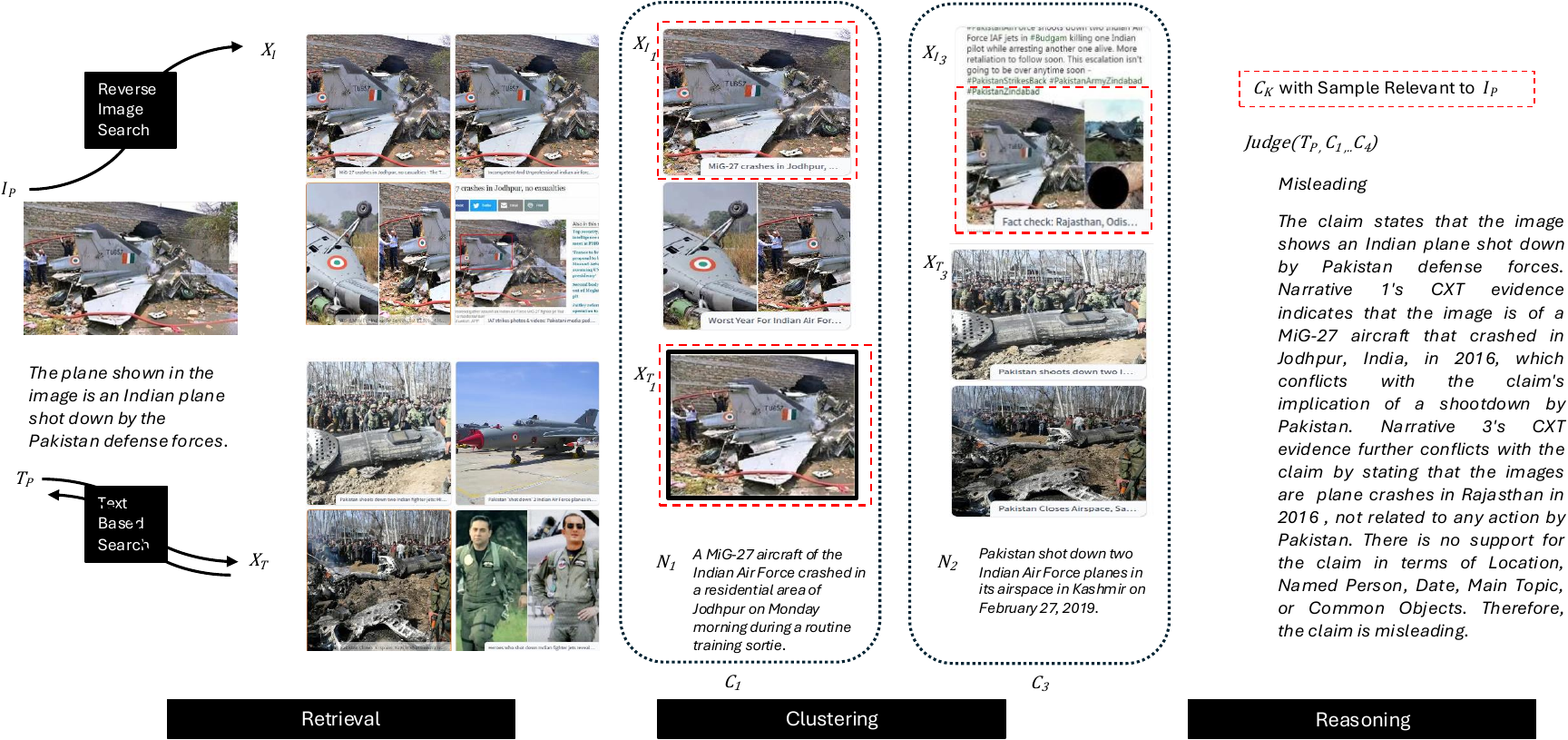}
    \caption{Overview of CRAVE: Given a claim (image + text), contained in a social media post, we retrieve evidence from reverse image and text-based searches. We then cluster the results into distinct narratives about the claim, selecting only visually relevant evidence (highlighted in red).  Each narrative is compared to the claim in a prompt-based LLM to predict veracity and produce a judgment and explanation. In this example the false claim—that the IAF plane was shot down by Pakistan—was countered by two clusters: C1 showing it was a training crash (true context), while C3 included unrelated real incidents of downed planes but crucially also a fact-checking article debunking the claim. This shows that the system can identify true context (C1) and legitimate debunks (C3).}
    \label{fig:main_frame}
\end{figure*}

LLM prompt-based veracity judgment has been  explored more recently, incorporating both example demonstrations and fine-tuning on training sets. Qi \etal~\citep{qi2024sniffer} predicts veracity by detecting inconsistencies between the caption, image, and visual entities using a fine-tuned Multimodal Large Language Model (MLLM)   InstructBLIP\citep{dai2023instructblipgeneralpurposevisionlanguagemodels}, and between the caption and text evidence using a frozen LLM \citep{vacunazheng2023judging}. The approach was extended to multiple debating LLMs in \cite{klakara2025llmconsensusmultiagentdebatevisual}. These approaches degrade when applied to images and entities with non-Western contexts that the MLLMs have not encountered during training. Thus~\citep{tonglet2025cove} uses predicted context and claim text as input, employing a frozen LLAMA 3~\citep{grattafiori2024llama} with few-shot demonstrations to predict veracity.  They also experiment with an LLM~\citep{he2021debertav3} fine-tuned on (predicted context, ground truth veracity) training pairs. However, the authors note that training on synthetic out-of-context data, which follow a different distribution than the real-world data,  worsens performance compared to direct comparison without learning. One insight from such work is that minimal learning may be more robust for final classification if one can structure how the evidence is compared. In CRAVE, we explore this and focus on structured claim-evidence comparison by combining narrative clustering and LLM judgment. The concept of judgment based on different narratives is explored in~\citep{khan2024debating}, by having expert models argue for different answers. However, multi-turn debates pose the risk of one side dominating and unduly influencing the decision.  In our approach, we simplify the process by using a single-step LLM judge that compares the generated narratives from the previous stage. This approach allows the judge to maintain greater control over the arguments considered, preventing any single viewpoint from influencing the outcome through prolonged interaction.

In summary, current research primarily focuses on debunking or verifying claims using ranked retrieved evidence. However, for divisive content, where different media outlets present conflicting narratives, fact-checking systems often fall short by failing to accommodate multiple perspectives. Community notes on X (formerly Twitter) illustrate how different user-sourced evidence can be used to support opposing views, leading to debate and engagement rather than a definitive verdict~\citep{de2024supernotes}. Arguably such an approach supports democratic discourse~\citep{walter2021unchecked}, but has not been leveraged in automated fact-checking work. CRAVE addresses this gap by using cluster-based, multimodal retrieval that can group evidence into coherent narratives, enabling LLMs to more effectively reason over long or conflicting inputs.

\section{\textbf{C}luster-based \textbf{R}etrieval \textbf{A}ugmented \textbf{V}erification with \textbf{E}xplanation}
The objective of CRAVE (\autoref{fig:main_frame}) is to predict and explain the veracity of a multimodal social media post \(P\), which comprises an image \(I_P\) and accompanying text \(T_P\), by leveraging external evidence. Our approach retrieves relevant evidence, organizes it into coherent narratives, and delivers a final veracity judgment with an explanation. The approach comprises three key stages: evidence retrieval, clustering, and  LLM reasoning.

\subsection{Retrieval}
The first step is to retrieve external evidence to contextualize the fact-checking process. 
 News articles are retrieved from fact-checking databases (e.g., Snopes, PolitiFact, Full Fact) and news archives from diverse sources (e.g., BBC, The Guardian, Reuters, Al Jazeera), enabling us to deal with both resurfaced and novel fake content. The system retrieves evidence using two complementary methods:

\paragraph{Reverse image search} 
    A reverse image search is conducted using the input image \(I_P\), retrieving a set of evidence pairs \(\mathcal{X}_I = \{(I_i, T_i)\}_{i=1}^{n_I}\), where each pair \((I_i, T_i)\) comprises an image \(I_i\) and its associated text \(T_i\). 
    Similar to~\citep{abdelnabi2022open,dey2024retrieval,tonglet2025cove}, we use the Vision API~\citep{google_vision_api} to perform reverse search. While~\citep{abdelnabi2022open} only uses the text entries from this reverse search, we use both returned evidence images and texts, similar to~\citep{dey2024retrieval}. This forms the image-based evidence set \(\mathcal{X}_I\) that represents the image context.  

\paragraph{Text-based search}
    The input text \(T_P\) is used to generate search queries with the assistance of an LLM which is applied to the Programmable search API~\citep{google_custom_search}. While the use of Programmable search has been seen in the past in~\citep{abdelnabi2022open,dey2024retrieval,tonglet2025cove}, we take forward the idea of generating search terms and iterative searches introduced in~\citep{dey2024retrieval}. These queries retrieve an additional set of pairs of evidence \(\mathcal{X}_T = \{(I_j, T_j)\}_{j=1}^{n_T}\), where each pair consists of an image \(I_j\) and an associated text \(T_j\). Our text-based search improves upon its initial retrievals through comparison of the evidence text with the claim text, to refine the search term for \textbf{H} subsequent iterations leveraging an LLM to directly discover inconsistencies. We determine the inconsistency in terms of extracted named entities related to 5W1H (Who, What, When, Where, Why, and How)~\citep{urbani2020verifying,tonglet2024image,tonglet2025cove}, with the search terms of the next iteration focused on entities not present in results from the current retrieval iteration. We try this for \textbf{H=3} iterations, stopping before if we cannot find discrepancies in terms of 5W1H. This process \textbf{augments} the evidence set. This refinement identifies missing 5W1H information in a evidence and generates targeted new queries to find it, adding more relevant evidence. This forms the text-based evidence set \(\mathcal{X}_T\).  

The combined evidence set is represented as:  
\[
\mathcal{X} = \mathcal{X}_I \cup \mathcal{X}_T.
\]

\subsection{Clustering: dealing with noisy evidence}
Web-scraped evidence is often noisy and may not directly relate to the claim. This issue is exacerbated in older datasets, where images are frequently repurposed in different contexts across platforms. For verbose social media posts, retrieving relevant evidence relies on extracting factual search terms from claims. Additionally, if a claim’s images and text do not originate from published news articles, retrieval becomes more challenging since such content is rarely indexed by search engines. This challenge is further aggravated when the post is not in English or does not concern prominent Western contexts. Prior work highlights these limitations, such as for Spanish tweets~\cite{dey2024retrieval} and image contextualization in East Africa and South Asia~\cite{tonglet2025cove}. To handle noisy evidence, we employ two main strategies: 
text-based clustering and visual filtering.   

\subsubsection{\textbf{Text-based clustering}}
Texts with different narratives coupled with irrelevant texts—whether due to scraping failures or website restrictions—can introduce noise during decision-making. To address this, we use narrative clustering to group texts about related topics into similar clusters. This helps maintain distinct narratives while preventing conflicted narratives from being mixed.

\paragraph{Initial clustering and narrative extraction}
Evidence clustering is one of our core contributions addressing several issues that arise in real-world fact-checking. First, web-scraped evidence is often noisy and contains irrelevant or contradictory information from diverse sources. As we later show, directly feeding all this evidence to an LLM can lead to lower accuracy or overly simplistic judgments. By grouping similar pieces of evidence into coherent `narratives', the LLM can focus on smaller, thematically consistent subsets of facts, mitigating the risk of overlooking an important detail. Second, clustering naturally separates conflicting claims into distinct clusters, enabling the system to reason about alternate viewpoints without conflating them. This is especially important for divisive content, where multiple sources may offer contradictory accounts of the same event.

The retrieved evidence set \(\mathcal{X}\) is clustered into \(K\) distinct groups using the K-means algorithm. Each piece of evidence from \(\mathcal{X}\) is first embedded into 384-dimensional dense vectors using Sentence Transformer~\cite{reimers-2019-sentence-bert}, applied to its text component \(T\). Our insights indicate that incorporating multimodal representations, such as image embeddings, could lead to degradation in classification performance due to the frequent pairing of identical images with conflicting text narratives. Each cluster \(\mathcal{C}_k\) is then defined as:
\[
\mathcal{C}_k = (\mathcal{X}_{I_k}, \mathcal{X}_{T_k})
\]
comprising of subset of evidence from reverse image search  \(\mathcal{X}_{I_k} \subseteq \mathcal{X}_I\)  and text based search  \(\mathcal{X}_{T_k} \subseteq \mathcal{X}_T\).   $\mathcal{X}_{I_k}$ or $\mathcal{X}_{T_k}$ can also be empty for a cluster, denoting the lack of any evidence from reverse image search or text based search respectively. We emphasize that  $\mathcal{X}_I$ and $\mathcal{X}_T$, and thus also $\mathcal{X}_{I_k}$ and $\mathcal{X}_{T_k}$,  are sets of evidence, with each evidence comprising of a text and an image component and thus homogeneous in representation, and this notation distinction is only so that we can prioritize reverse search evidence during judgment.

Finally, a narrative \(N_k\) is selected for each cluster by taking the text component of the most representative element within \(\mathcal{C}_{k}\), measured as its distance from the centroid, yielding the final cluster representation:
\[
\mathcal{C}_k = (\mathcal{X}_{I_k}, \mathcal{X}_{T_k}, N_k)
\]
The choice of \(K=4\) was determined through experimentation on Real-world fake news benchmark Verite, 5pilsOOC, and DP, as analyzed in \autoref{subsec:clusterchoicek}, while the clustering method was determined through experimentation on random subsets of 300 samples each from these datasets, analyzed in \autoref{subsec:clusterchoicemethod}. These coherent clusters not only allow us to deal with noisy web data during judgment but also enable retrieval of targeted evidence detailed in \autoref{subsec:agent}.

\paragraph{Agent-based cluster refinement}\label{subsec:agent}
An instance of text-based search is applied, with \textbf{H=2},  to each cluster \(\mathcal{C}_k\) to obtain cluster-specific evidence through the following targeted evidence retrieval. Additional queries are generated based on inconsistencies with the narrative \(N_k\) to augment the text-based evidence, forming a refined set \(\mathcal{X}_{T_k}^\prime\).    
The refined cluster representation becomes:  
\[
\mathcal{C}_k^\prime = (\mathcal{X}_{I_k}, \mathcal{X}_{T_k}^\prime, N_k).
\]

\subsubsection{\textbf{Visual filtering}}  
Low visual similarity reliably indicates irrelevant evidence, thus we filter evidence using an image similarity threshold, following prior approaches~\cite{tonglet2025cove,dey2024retrieval}. Our visual features are extracted from pre-trained networks that capture key visual aspects: faces, scenes, and visual semantics. Faces are detected using~\cite{mtcnnzhang2016joint}, with the most prominent face embedded via~\cite{ct_schroff2015facenet}, yielding $\mathcal{V}^{face} \in \mathbb{R}^{512}$. Scene information is extracted via a model trained on 365 place categories \cite{zhou2017places}, producing $\mathcal{V}^{place} \in \mathbb{R}^{2048}$. Visual semantics are captured using~\cite{vit_dosovitskiy2020image}, resulting in $\mathcal{V}^{sem} \in \mathbb{R}^{1000}$. The final visual representation of an image is:  

\[
\mathcal{V} = [ \mathcal{V}^{face}, \mathcal{V}^{place}, \mathcal{V}^{sem} ].
\]

The cosine similarity between $\mathcal{V}^{I_P}$ and  $\mathcal{V}^{I_{k}}$  defines the final similarity score of an evidence image  \(I_k\) $\in$ $\mathcal{C}_k^\prime$   with the claim  image \(I_P\). We compute cosine similarity for each vector type (face, place, semantics) separately and then combine with equal weighting, avoiding dimensionality issues. Following~\cite{tonglet2025cove} we leverage the large training set of NewsCLIPpings to determine our similarity threshold of 0.9. 

Since reverse image searches are designed to retrieve high-quality, contextually relevant images, evidence with low visual similarity—typically originating from text-based searches—can be safely disregarded, as it is unlikely to offer meaningful context for verification. However, visually similar yet textually different evidence may indicate out-of-context usage rather than irrelevance. Thus by using visual similarity for filtering and clustering noisy textual evidence, we ensure that relevant evidence, even if with a different text context, is retained and presented in a structured manner to the LLM for reasoning.

\begin{figure}[h]
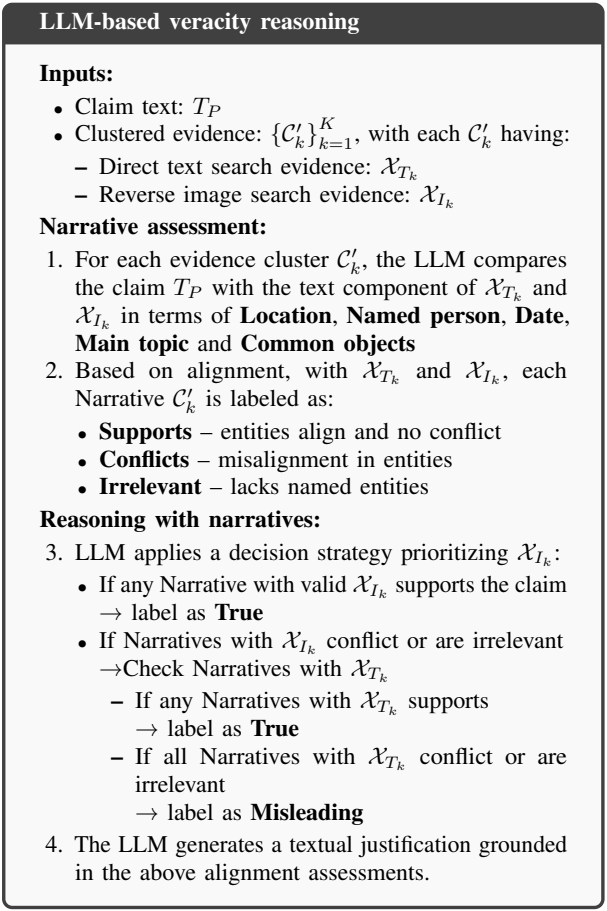

\centering
\scalebox{0.9}{
\begin{tcolorbox}[colback=gray!5!white,colframe=black!75!white,
title={LLM-based veracity reasoning}, fonttitle=\bfseries]
\textbf{Inputs:}
\begin{itemize}[leftmargin=1.5em]
    \item Claim text: $T_P$
    \item Clustered evidence: $\left\{ \mathcal{C}_k^\prime \right\}_{k=1}^{K}$, with each  $\mathcal{C}_k^\prime$ having:
    \begin{itemize}
        \item Direct text search evidence: $\mathcal{X}_{T_k}$
        \item Reverse image search evidence: $\mathcal{X}_{I_k}$
    \end{itemize}
\end{itemize}
\textbf{Narrative assessment:}
\begin{enumerate}[leftmargin=1.5em, label*=\arabic*.]
    \item For each evidence cluster $\mathcal{C}_k^\prime$, the LLM compares the claim $T_P$ with the text component of $\mathcal{X}_{T_k}$ and $\mathcal{X}_{I_k}$ in terms of   \textbf{Location}, \textbf{Named person}, \textbf{Date}, \textbf{Main topic} and \textbf{Common objects}

    \item Based on alignment, with $\mathcal{X}_{T_k}$ and $\mathcal{X}_{I_k}$, each Narrative $\mathcal{C}_k^\prime$ is labeled as:
    \begin{itemize}
        \item \textbf{Supports} – entities align and no conflict
        \item \textbf{Conflicts} – misalignment in entities
        \item \textbf{Irrelevant} – lacks named entities
    \end{itemize}
\end{enumerate}
\textbf{Reasoning with narratives:}
\begin{enumerate}[resume, leftmargin=1.5em, label*=\arabic*.]
    \item LLM applies a decision strategy prioritizing $\mathcal{X}_{I_k}$:
    \begin{itemize}
        \item If any Narrative with valid $\mathcal{X}_{I_k}$ supports the claim \\
           → label as \textbf{True}
        \item If Narratives with $\mathcal{X}_{I_k}$ conflict or are irrelevant\\
              →Check Narratives with $\mathcal{X}_{T_k}$
             \begin{itemize}
                \item If any Narratives with $\mathcal{X}_{T_k}$ supports\\
                 → label as \textbf{True}
                \item If all Narratives with $\mathcal{X}_{T_k}$ conflict or are irrelevant\\
                 → label as \textbf{Misleading}
              \end{itemize}
    \end{itemize}
    \item The LLM generates a textual justification grounded in the above alignment assessments.
\end{enumerate}
\end{tcolorbox}
}
\caption{Reasoning protocol used by the LLM (GPT 4o) to assess the veracity of a post $(T_P, I_P)$ using clustered direct ($\mathcal{X}_{T_k}$) and reverse image search ($\mathcal{X}_{I_k}$) evidence. Narrative assessment includes comparison and labeling, while reasoning prioritizes reverse image evidence to determine a final verdict.}
\label{fig:llm-veracity-protocol}
\end{figure}

\subsection{Reasoning: prompt-based comparison and explanation }\label{subsec:Judgement}

The refined clusters \(\{\mathcal{C}_k^\prime\}_{k=1}^{K}\) and their associated narratives \(\{N_k\}_{k=1}^{K}\) are evaluated by an LLM-based judge along with the original social media post \(P\), to produce a veracity verdict and explanation.

The verdict categories are:  
\begin{itemize}
    \item \textbf{True:} The post is factually accurate.
    \item \textbf{Misleading:} The post is factually inaccurate.
    \item \textbf{Not enough data:} Insufficient evidence to determine the veracity of the post.
\end{itemize}
Our Judgment Phase comprises comparison with the Claim leading to an interpretable explanation and coherent verdict.  Our prompt, \autoref{fig:llm-veracity-protocol}, evaluates the veracity of a social media post \( P = (T_P, I_P) \), consisting of a textual claim \( T_P \) and an associated image \( I_P \), by leveraging clustered evidence obtained from both direct and reverse image search. Each narrative cluster \( \mathcal{C}_k^\prime \) contains evidence from text-based search (\( \mathcal{X}_{T_k} \)) and evidence from reverse search (\( \mathcal{X}_{I_k} \)). 

\subsubsection{Narrative assessment} The judgment process begins with narrative assessment, where the LLM compares the textual content of each evidence item in \( \mathcal{C}_k^\prime \) against the claim text \( T_P \). Unlike prior methods that treat context prediction as a standalone step~\citep{tonglet2025cove}, our method integrates contextual understanding directly into a unified prompting strategy. Using a 5W1H-inspired questioning framework—focusing on who, what, when, where, why, and how—the LLM evaluates alignment between the claim and evidence across five critical dimensions: named entities, dates, locations, key topics, and common objects. This entity-level comparison helps determine whether an evidence item supports the claim, conflicts with it, or lacks relevant content entirely. 
In real-world scenarios, each narrative cluster often includes a mixture of evidence with varying degrees of fine-grained support. Such variability introduces noise into the reasoning process, making it challenging to draw accurate conclusions if relying solely on the highest-ranked evidence within a narrative. To address this, we aggregate evidence holistically at the cluster level, treating each \( \mathcal{C}_k^\prime \) as a coherent narrative. Rather than favoring individual pieces of evidence in isolation, the LLM considers the collective alignment pattern within each cluster. The output of this stage is a structured set of alignment labels for each narrative, guiding further reasoning.

\subsubsection{Reasoning with narratives} Following narrative assessment, where each narrative cluster \( \mathcal{C}_k^\prime \) is labeled based on entity-level alignment, the LLM proceeds to reason over these narratives to derive a final veracity decision. This reasoning process is grounded in the principle that narrative coherence, especially when supported by image provenance, is critical for verifying factual claims. Specifically, the LLM first identifies whether there exists any narrative cluster that both (i) contains valid reverse image search evidence (\( \mathcal{X}_{I_k} \)), and (ii) has been labeled as \textbf{Supports} during narrative assessment. If such a narrative is found, the post is immediately labeled as \textbf{True}. If no supporting narrative with valid reverse evidence is available, the model then checks for supporting narratives that contain only text-based evidence (\( \mathcal{X}_{T_k} \)). If at least one such narrative supports the claim, the label remains \textbf{True}. However, if all narratives—whether supported by image or text searches—are either labeled as \textbf{Conflicts} or \textbf{Irrelevant}, the post is ultimately labeled as \textbf{Misleading}. 
This structured approach ensures that multiple sources of evidence are considered in a tiered fashion, reducing the risk of false negatives or misclassifications due to noisy or incomplete top-ranked results. Finally, the LLM produces a natural language explanation grounded in the alignment rationale, clearly articulating the justification behind the final verdict and enhancing the transparency and interpretability of the model’s reasoning. To ensure transparency and facilitate the reproduction of our results, the code and data supporting this study will be made publicly available upon acceptance.

\section{Datasets}
\subsection{Current datasets }\label{subsec:cur_ds}
Current misinformation datasets present several limitations~\citep{shahid2022detecting} that hinder the comprehensive analysis of real-world misinformation on social media. Many are synthetically generated~\citep{liu2024mmfakebench,luo2021newsclippings,abdelnabi2022open}, which, while useful for specific evaluations, often fail to capture the nuanced and evolving nature of online misinformation.

In~\citep{luo2021newsclippings}, some true samples used generic file images unrelated to the claim text, with their truth determined by the context in which news websites used these images. Evidence collected by~\citep{abdelnabi2022open} at the time aligned with this context, but many of these images have since been repurposed, causing misalignment in reverse image search evidence. Moreover, these datasets may not adequately reflect the linguistic complexities, contextual variations, and multimodal characteristics of real-world fake news. 

Thus it is important to focus on real-world fake news datasets as introduced in~\citep{papadopoulos2024verite,tonglet2025cove}. However, these datasets contain curated and sanitized evidence, which bypasses the significant challenge of retrieving relevant information from the vast and noisy web in practical scenarios. This is particularly problematic when dealing with divisive content, where different media outlets and individuals may present conflicting narratives and supporting evidence, a phenomenon often overlooked by traditional fact-checking models that tend to seek a singular truth. Consequently, there is a pressing need for datasets that better represent the complexities of real-world fake news, especially those surrounding topics where differing opinions and diverse sources contribute to a fragmented information landscape. 
To address these gaps, we construct a dataset that captures the complexity of online discourse—particularly the challenge of verifying claims amid multiple, and sometimes conflicting, perspectives and evidence.

\begin{table}[H]
    \centering
    \begin{tabular}{l|c}
        \hline
        \textbf{Characteristic} & \textbf{Count} \\
        \hline
        Total tweets & 608 \\
        Labels: \texttt{True} & 359  \\
        Labels: \texttt{Misleading} & 225 \\
        Labels: \texttt{Conflicted} & 8  \\ \hline
        Average notes per tweet & 2.54\\ 
        Average ratings per note & 138 \\
        Notes labeled as "helpful" & 0.59 \\
        Notes labeled as "unhelpful" & 0.37 \\ \hline
    \end{tabular}
    \caption{Statistical overview of the DP dataset.}
    \label{tab:dataset_stats}
\end{table}

\subsection{Divisive Posts (\textbf{DP}) evaluation benchmark}
We constructed our multimodal dataset from X (formerly Twitter) Community Notes, a collaborative fact-checking system where contributors identify potentially misleading information. Our collection process followed a systematic approach to ensure quality and relevance. We systematically curated our dataset by first collecting daily Community Notes and their corresponding ratings from X’s transparency initiative. We then performed frequency analysis on note records, prioritizing tweets with the highest number of notes to target content with high engagement and potential controversy. For each selected tweet, we integrated ratings by mapping and filtering to the top 5 rated notes, ensuring quality and diverse perspectives. Using the X API, we retrieved the full tweet content, including text and media, to support multimodal analysis. Only tweets containing both text and at least one image were retained. Finally, human annotators manually labeled each tweet, referencing external sources to establish ground truth veracity. The resulting dataset, details in  \autoref{tab:dataset_stats}, provides a benchmark for evaluating multimodal misinformation detection models, as it captures real-world instances of potentially misleading content along with community-generated assessments. While Community Notes data is publicly available, we ensured ethical usage by removing personal identifiers, focusing only on public account content, and fully complying with X’s developer terms of service.

\section{Experiments and results}

\begin{table*}[t]
\centering
\begin{tabular}{lccccccc}
\hline
\textbf{Method}                                                      &  \multicolumn{3}{c}{\textbf{Verite}}   & \textbf{MMFakeBench}  & \textbf{5PilsOOC}        & \textbf{NewsCLIPpings}    & \textbf{DP}\\ \hline
                                                                     & T/F  & T/O   & T/M                     & T/F                   & T/O                       & T/O                       & T/F         \\
MMbase~\citep{liu2024mmfakebench}                         &  -   &   -   &  -                      & \textit{0.76}                 &   -                       &    -                      &  -          \\         
 
AITR~\citep{papadopoulos2024similarity}                              & 0.52(\textit{(0.55}) & 0.73(\textit{0.80}) & 0.52(\textit{0.49})   &  0.59                 & 0.49 (\textit{0.52})                    & \textbf{\textit{0.90}}                      &  0.64       \\

MUSE~\citep{papadopoulos2024similarity}                              & 0.53(\textit{0.57}) &0.74( \textit{0.80})  &0.52(\textit{0.51})   &  0.58                 & 0.49                      & \textit{0.89}                      &  0.63       \\ 
RED-DOT~\citep{papadopoulos2024reddotmultimodalfactcheckingrelevant} &  -   & \textit{0.76}  &  -                      &  0.60                    & \textit{0.46}                      & \textit{0.88}                      & 0.61      \\ 
COVE~\cite{tonglet2025cove}                                          &  -   &   -   &  -                      &  -                    & \textit{0.58}                      & \textit{0.86}                      &  -          \\  
CRAVE                                                                 &\textbf{0.82} &\textbf{0.80}  & \textbf{0.80}                    & \textbf{0.78}                  &\textbf{0.81}                     & \textit{0.85}                      &  \textbf{0.73} (\textit{0.64})       \\ \hline
\end{tabular}
\caption{ Accuracy comparison against  state-of-the-art methods, across datasets. T/O : True versus Out-Of-Context(\textbf{OOC}), T/F : True versus Fake, T/M: True versus Miscaptioned. For Verite T/F, we consider both  OOC and Miscaptioned to be fake. For Verite and 5PilsOOC we report the accuracy using the evidence that came with the datasets in  italics. For NewsCLIPpings we don't collect evidence and use the evidence that came with the dataset. For DP, the numbers in parentheses indicate the accuracy only using the community notes as evidence.}
\label{tab:sota}
\end{table*}
\subsection{Comparison against state-of-the-art methods}

We focus our efforts on real-world fake news datasets where
evidence retrieval and refinement can be more effectively applied to improve performance. \autoref{tab:sota} demonstrates that the binary veracity prediction results of our zero-shot model CRAVE improve upon the state-of-the-art results on both real-world datasets Verite and 5PilsOOC, as well as the Synthetic dataset MMFakeBench. For NewsCLIPpings, due to the issues outlined in \autoref{subsec:cur_ds}, we refrain from evidence collection or refinement, instead use the evidence provided~\citep{abdelnabi2022open}. We apply an ablated version of CRAVE that uses only the LLM reasoning step, achieving competitive results with the zero-shot model COVE, though not matching the performance of fully trained models. The table also shows that  the performance of these trained models degrade when dealing with our noisy web-scraped evidence data versus the curated and cleaned evidence data that came with the datasets(denoted in parentheses). While these methods use different pipelines for evidence collection, we use a single evidence collection pipeline, refined through clustering and filtering. For DP, we also test against the Community notes data to validate the superior quality of our collected evidence.

\subsection{Baseline performance using MLLMs}
\autoref{tab:llm} compares Multimodal Large Language Models (MLLMs) with our CRAVE framework on the binary veracity prediction task. For both InstructBLIP~\citep{dai2023instructblipgeneralpurposevisionlanguagemodels} and LLAVA~\citep{llavaliu2024improvedbaselinesvisualinstruction}, we use Vicuna-13B~\citep{vacunazheng2023judging} as the underlying language model. All models are evaluated on the same input: the claim and the retrieved evidence. However, performance varies significantly due to how they process and reason over noisy, multimodal contexts.

We report results for three configurations: \textbf{C} (claim-only), \textbf{V} (claim with both images and text as evidence), and \textbf{T} (claim with text-only evidence). Since LLAVA cannot process multiple images at once, we report only the \textbf{T} configuration for LLAVA, using the text component of the evidence.

\begin{table}[H]
\scriptsize
\setlength{\tabcolsep}{3pt} 
\centering
    \begin{tabular}{lc|c|c|c|c}
        \toprule
        \textbf{Method}      & \textbf{Verite} & \textbf{MMFakebench} & \textbf{5PilsOOC} & \textbf{NewsCLIPpings} & \textbf{DP} \\
                             & Acc & Acc & Acc  & Acc  & Acc \\
        \midrule
        InstructBLIP (C)      & 0.50       & 0.50          & 0.55     & 0.55             & 0.54         \\  
        InstructBLIP (V)      & 0.63       & 0.62          & 0.50     & 0.50             & 0.40         \\ 
        InstructBLIP (T)      & 0.56       & 0.50          & 0.50     & 0.47             & 0.43          \\ \hline  
        LLAVA (C)             & 0.66       & 0.68          & 0.53     & 0.54             & 0.40          \\
        LLAVA (T)             & 0.68       & 0.71          & 0.60     & 0.69             & 0.49          \\\hline  
        GPT 4 (V)            & 0.78~\citep{braun2025defamedynamicevidencebasedfactchecking}       & 0.75~\citep{liu2024mmfakebench}          & -        &   -             & -          \\\hline  
        CRAVE                & 0.82       & 0.78          & 0.81     & 0.85             & 0.73         \\
        \bottomrule
    \end{tabular}
\caption{Comparison against  MLLMs across datasets.   C denotes models using only the Claim. V denotes the version using both images and text as evidence, and T denotes versions using only the text component of the evidence. }
\label{tab:llm}
\end{table}

When relying solely on the claim (denoted as \textbf{C}), MLLMs exhibit limited predictive power, reflecting their restricted commonsense or factual grounding in the absence of external evidence. This is evident from the relatively low performance of both InstructBLIP and LLAVA under this setting. Upon incorporating retrieved evidence, the behavior of the models varies considerably. LLAVA, using text-based evidence (\textbf{T}), consistently improves. In contrast, InstructBLIP shows more volatility, with performance degrading in some datasets, particularly when using image-based evidence. This highlights the challenges MLLMs face when integrating noisy or conflicting multimodal evidence.

A key insight is the stronger role of language in reasoning. Both MLLMs perform better with text-based evidence (\textbf{T}) than with image-based evidence (\textbf{V}). This suggests that language provides clearer cues for claim verification, especially in misinformation contexts where images may be reused or repurposed across narratives. This aligns with CRAVE’s design, which relies on language for decision-making, using visual cues only to filter evidence.

CRAVE’s superior performance stems from its ability to structure and disambiguate noisy evidence. Rather than reasoning over large, unfiltered contexts, CRAVE organizes evidence into coherent narrative clusters, enabling it to evaluate claims against dominant narratives instead of isolated fragments. This structure-aware approach minimizes confusion from conflicting sources and focuses decision-making on linguistically consistent, relevant information.

Additionally, unlike MLLMs—whose performance may be skewed by biases from pretraining on large-scale web data—CRAVE is less influenced by domain priors or cultural biases. This is particularly advantageous in datasets like 5PilsOOC and DP, where MLLMs struggle with claims from underrepresented regions or topics. Finally, standard prompting – asking a powerful model to fact-check an image/text pair – often fails to deal with conflicting noisy evidence. CRAVE’s zero-shot approach and reliance on structured narratives lead to more robust decision-making, regardless of cultural or regional context.

In summary, while MLLMs show potential with textual evidence (\textbf{T}), their lack of structural reasoning and sensitivity to noisy evidence limit their utility for veracity prediction. CRAVE overcomes these challenges through structured narrative alignment and language-based reasoning, providing more robust and generalizable performance across diverse misinformation scenarios.


\subsection{Ablation study}
To validate our design choices and quantify the contribution of each module, we conduct a comprehensive ablation study. We first dissect the impact of our framework's core architectural components, such as iterative refinement and narrative clustering. We assess the system's sensitivity to external factors, evaluating various backbone LLMs and evidence modalities to understand their influence on overall performance.
Finally, we then delve into the clustering process itself, analyzing the optimal cluster count and the choice of method.

\subsubsection{Study 1 - Role of components}
\autoref{tab:refine} demonstrates the effectiveness of iterative refinement during text evidence $\mathcal{X}_{T}$ retrieval which leads to improved scores across datasets. 
Clustering of all evidence into distinct narratives leads to improved veracity prediction compared to taking the maximally similar evidence. Finally Agentic Retrieval of $\mathcal{X}_{T_k}$ based on $K$ individual narrative clusters leads to further improvements for all datasets except Verite; this could be attributed to the lack of novel evidence retrieval that is already not part of the evidence set.

\begin{table}[h]
\scriptsize
\setlength{\tabcolsep}{1pt} %
    \centering
    \begin{tabular}{lllcc|cc|cc|cc|cc}
        \toprule
        $\mathcal{X}_{T}$  & Cluster   &  $\mathcal{X}_{T_k}$                         & \multicolumn{2}{c}{\textbf{ Verite }} & \multicolumn{2}{c}{\textbf{MMFakebench}} & \multicolumn{2}{c}{\textbf{5PilsOOC}} & \multicolumn{2}{c}{\textbf{NewsCLIPpings}} & \multicolumn{2}{c}{\textbf{DP}} \\
         Refine & Judge  & Refine  & Acc & F1 & Acc & F1 & Acc & F1 & Acc & F1 & Acc & F1\\
        \midrule
        $\times$      & $\times$      & $\times$        & 0.75     & 0.74     & 0.71         & 0.68        & -        & -     & 0.65     & 0.65  & -        &  -    \\
        $\times$      & $\checkmark$  & $\times$        & 0.82     & 0.81     & 0.75         & 0.71        & -        & -     & 0.85     & 0.85  & -        &  -    \\
        $\checkmark$  & $\times$      & $\times$        & 0.74     & 0.73     & 0.72         & 0.71        &0.72      &0.71   & -        &  -    & 0.62     & 0.61  \\
        $\checkmark$  & $\checkmark$  & $\times$ &\textbf{0.84}    & 0.83     & 0.76         & 0.74        &0.81      &0.81   & -        &  -    & 0.73     & 0.73  \\
        $\checkmark$  & $\checkmark$  & $\checkmark$    & 0.82     & 0.82   &\textbf{0.78}   & 0.76  &\textbf{0.81}   &0.81   & -        &  -    &\textbf{0.73}     & 0.73  \\
        \bottomrule
    \end{tabular}
    \caption{ Role of components: Improvements due to text evidence refinement ( $\mathcal{X}_{T}$ Refine), Clustering before judgment(Cluster Judge) and Evidence refinement per cluster ( $\mathcal{X}_{T_k}$ Refine).}
\label{tab:refine}
\end{table}

\begin{figure*}[htbp]
    \centering
    \includegraphics[width=0.9\textwidth]{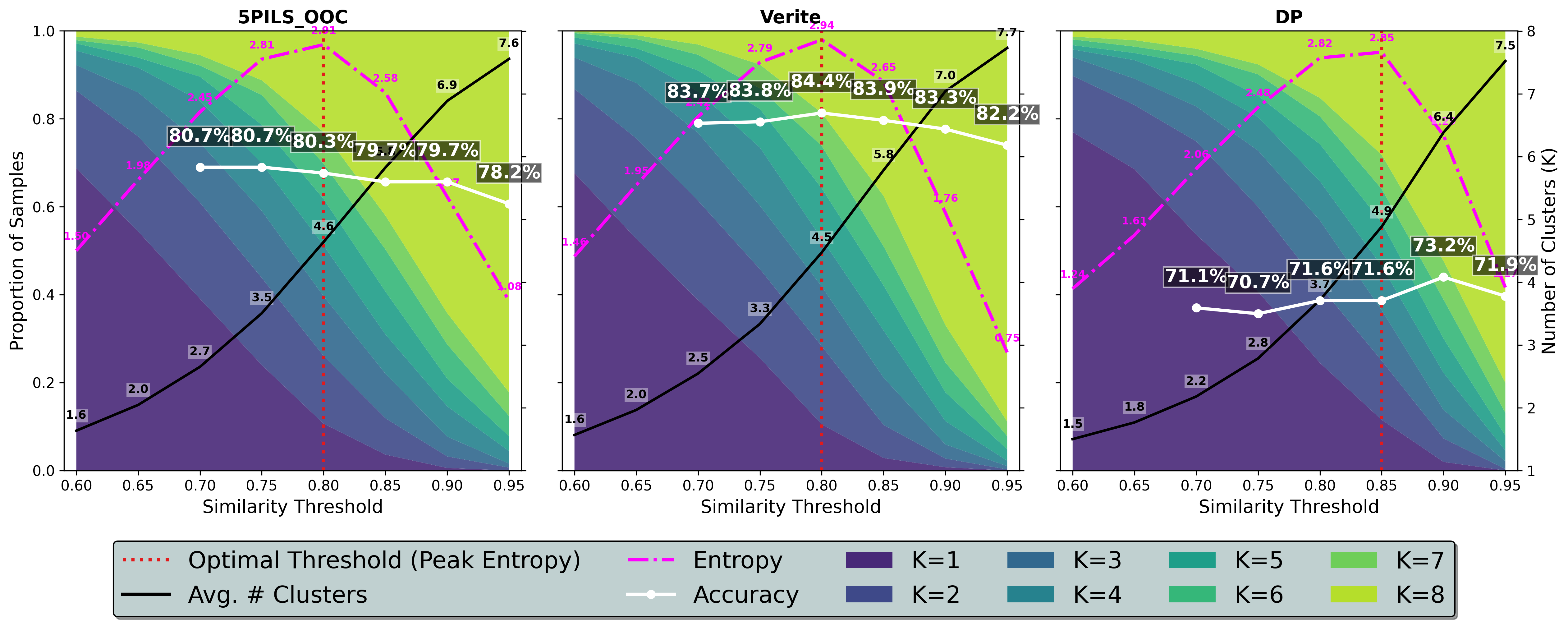}
    \caption{Dynamic Clustering performance varies with dataset complexity, as higher similarity thresholds fragment evidence into more clusters. At peak entropy—where clustering uncertainty is highest—samples span 4 clusters on average. Since the optimal threshold is claim-dependent and cannot be determined a priori, we adopt a static clustering with K=4.}
    \label{fig:clusterscheme}
\end{figure*}

\subsubsection{Study 2 - Backbone LLM and evidence stream}

\begin{table}[htbp]
\centering
\setlength{\tabcolsep}{4pt} %
\begin{tabular}{lccc}
\toprule
\textbf{Backbone LLM for Judgment} & \textbf{Verite} & \textbf{5PilsOOC} & \textbf{DP} \\
\midrule
GPT-4o & 0.82 & 0.81 & 0.73 \\
Gemini-2.0-Flash & 0.77 & 0.83 & 0.74 \\
Llama-3.1-8B-Instruct & 0.77 & 0.77 & 0.67 \\ \hline
Llama-3.1-8B-Instruct (I) & 0.75 & 0.75 & 0.69 \\
Llama-3.1-8B-Instruct (T) & 0.72 & 0.73 & 0.66 \\
\bottomrule
\end{tabular}
\caption{Backbone LLM and evidence Stream. Commercial LLMs excel  due to superior reasoning; reverse image search(I) outperforms text-only(T), with T+I combination being optimal}\label{tab:llm_veracity_comparison}
\end{table}

In \autoref{tab:llm_veracity_comparison}  we study various backbone LLMs for veracity prediction accuracy. Commercial models perform on par, with a notable boost on the complex DP dataset, likely due to superior reasoning. However, even with open models like Llama we improve upon the current state of the art.  Unsurprisingly image provenance through Reverse image search proves to be a stronger signal that text search (T) alone. While combining them (T+I)  usually leads to better performance, on the divisive DP dataset it often gets overwhelmed by conflicting narratives from the text evidence.

\subsubsection{Study 3 - Cluster size}\label{subsec:clusterchoicek}
Our analysis of Dynamic(Agglomerative) Clustering performance, \autoref{fig:clusterscheme}, reveals that the optimal configuration is intrinsically tied to the complexity of the underlying dataset. We identify the optimal similarity threshold, by locating the peak of the clustering entropy—a point representing maximum algorithmic uncertainty and a critical transitional state. For datasets like 5PILSOOC and Verite, curated from fact-checking corpora, this heuristic aligns well with peak classification accuracy. 5PILSOOC, with its verbatim claims, achieves effective classification at a low threshold (0.70) requiring an average of only 2.7 clusters. Verite, featuring paraphrased claims, introduces greater semantic ambiguity, thus demanding a higher threshold (0.80) and a larger average cluster count (K=4.5) to resolve these nuances.

In stark contrast, our DP dataset, derived from informal Community Notes, presents a more complex challenge. Here, peak accuracy ($73.2\%$) is achieved only at a stringent threshold of 0.90, well after the point of peak entropy (which occurs at 0.85). This divergence demonstrates that for nuanced real-world fake claims, the system must be pushed beyond its state of maximum uncertainty, requiring an average of 6.4 clusters to isolate the correct classifying signal from a large set of distractors.

This variance underscores a critical insight: an optimal dynamic clustering threshold is claim-dependent and cannot be determined a priori for a general-purpose system. 
Lacking a reliable method to predict claim complexity for an adaptive threshold, we adopt a static clustering with K=4, a value empirically grounded as the average number of clusters per sample (4.0) observed at peak entropy across our datasets.

\begin{table}[h]
\scriptsize
\setlength{\tabcolsep}{3pt} %
    \centering
    \begin{tabular}{l | ccc | ccc | ccc}
        \toprule
        \textbf{Method}  & \multicolumn{3}{c|}{5PilsOOC} & \multicolumn{3}{c|}{Verite} & \multicolumn{3}{c}{DP} \\ \hline
                                      & Sil ↑ & DB  ↓ & Acc ↑ & Sil ↑ & DB ↓ & Acc ↑ & Sil ↑ & DB ↓ & Acc ↑\\

        Kmeans                       & 0.32  & 1.19  &  0.81      & 0.28  & 1.26  & 0.80     & 0.30  & 1.18 & 0.70    \\
        GPT 4o-mini                   & 0.35  & 1.08  &  0.78      & 0.33  & 1.12  & 0.79     & 0.31  & 1.21 & 0.67   \\    
        Spectral                     & 0.19  & 1.47  &  0.74      & 0.18  & 1.50  & 0.72     & 0.19  & 1.46 & 0.66   \\
 
        \bottomrule
    \end{tabular}
    \caption{Cluster coherence metrics for different method. For all datasets, k-means consistently outperforms spectral clustering.}
    \label{tab:clustering_metrics}

\end{table}

\subsubsection{Study 4 - Cluster method}\label{subsec:clusterchoicemethod}
The clustering performance metrics, in \autoref{tab:clustering_metrics}, reveal key insights into the effectiveness of different methods across random subsets (300 samples each) of the real-world datasets 5PilsOOC, Verite, and DP. The evaluation metrics include the  average silhouette score (Sil)~\citep{rousseeuw1987silhouettes}  (indicating cluster cohesion and separation, with higher values being better) and the Davies-Bouldin(DB)~\citep{davies1979cluster} index (where lower values indicate better-defined clusters) along with the veracity accuracy. We also compare against a GPT 4o-mini prompt-based clustering. While the cluster coherence was competitive, if not better at times,  GPT 4o-mini often excluded  outlier evidence from the clusters.

Overall, K-means clustering consistently outperforms spectral clustering, achieving the higher silhouette scores and lowest Davies-Bouldin indices across all datasets, indicating well-defined and compact clusters. In contrast, spectral clustering struggles suggesting less optimal separation. Dataset-specific trends also emerge, with 5PilsOOC exhibiting the best clustering structure, while Verite appears to be the most challenging, as indicated by its lower silhouette scores, especially for spectral clustering. Interestingly, GPT 4o-mini forms the most coherent clusters, but this is often at the cost of completely ignoring outliers which leads to the lower overall accuracy.

\subsection{User study: Explanation quality and preference }

We evaluate the quality of and preferences for CRAVE-generated explanations through user studies. The evaluation is divided into two studies. In the first study, we assess the clarity and comprehensiveness of the generated explanations, while in the second study, we compare the preference for the generated explanation over community notes.

\subsubsection{Study 1 - Clarity and comprehensiveness evaluation}

To assess the quality of fact-checking explanations produced by our system, we conducted a human evaluation using Amazon Mechanical Turk \textbf{(AMT)}. Participants were shown a news story along with a set of retrieved narrative facts, the system-generated veracity prediction, and the system-generated explanation. Their task was to rate the explanation on two dimensions: clarity and comprehensiveness. Clarity refers to how easy the explanation is to understand, and comprehensiveness measures the extent to which the explanation incorporates key facts to justify the verdict. Ratings were provided on a 5-point Likert scale (using discrete values 1, 3, and 5).  AMT workers were filtered to include only those who spent more than one minute on the task and had a historical approval rate above 95\%; out of 56 participants, 39 met these criteria. \autoref{tab:eval_stats} summarizes the evaluation statistics for both LLM-generated scores and the filtered AMT evaluations. While the participants were not aware of the ground truth, we partitioned the scores for correct and incorrect predictions. The data indicate that explanations for correct predictions generally receive higher clarity scores than those under incorrect predictions. For example, LLM evaluations show an average clarity score of 3.28 for correct predictions versus 2.99 for incorrect ones. AMT evaluations yield higher scores overall, underscoring the value of human judgment in assessing explanation quality to detect nuances in explanation quality more reliably.

\begin{table}[h]
\centering
\scriptsize
\setlength{\tabcolsep}{3pt} 
\begin{tabular}{l|c|cccc|cccc}
\toprule
\textbf{Source / Condition} & \textbf{Workers} & \multicolumn{4}{c|}{\textbf{Comprehensiveness}} & \multicolumn{4}{c}{\textbf{Clarity}} \\
 &  & \textbf{1} & \textbf{3} & \textbf{5} & \textbf{Avg} & \textbf{1} & \textbf{3} & \textbf{5} & \textbf{Avg} \\
\midrule
\textbf{LLM Overall}         & N/A  & 99  & 387 & 122 & 3.08   & 47  & 455 & 106 & 3.19  \\
LLM Correct         & N/A  & 47  & 293 & 94  & 3.22   & 16  & 342 & 76  & 3.28  \\
LLM Incorrect       & N/A  & 52  & 94  & 28  & 2.72   & 31  & 113 & 30  & 2.99  \\
\midrule
\textbf{AMT Overall} & 39   & 156 & 872 & 586 & 3.53   & 162 & 780 & 672 & 3.63  \\
AMT  Correct & 36   & 115 & 628 & 425 & 3.53   & 111 & 569 & 488 & 3.65  \\
AMT  Incorrect & 24  & 41  & 244 & 161 & 3.54   & 51  & 211 & 184 & 3.60  \\

\bottomrule
\end{tabular}
\caption{Evaluation statistics for CRAVE explanation quality. For each participant (LLM, AMT) and partition (Overall, Correct predictions, Incorrect predictions), the table reports counts for ratings along with the average score for both comprehensiveness and clarity.}
\label{tab:eval_stats}
\end{table}

\subsubsection{Study 2 - Preference comparison of explanations}
In the second study, we examine the preference between two types of explanations for a news story: the user-generated explanation (community notes) and the system-generated explanation. The task is to select the explanation that is more intuitive, clear, and concise, and that better supports the story’s ground truth veracity. \autoref{tab:preference_comparison} presents the preference fractions from both LLM  and AMT workers. While LLM assessments show a preference of 67.11\% for the CRAVE-generated explanation compared to 32.89\% for the user explanation, AMT workers—based on a sample of 10 workers—exhibited a preference of 57.66\% for the CRAVE explanation versus 42.34\% for the user-generated notes. These findings suggest that, across both evaluation modalities, the system-generated explanation is consistently viewed as more effective in conveying the veracity of the story.
\begin{table}[h]
\centering
\begin{tabular}{lccc}
\toprule
\textbf{Source} & \multicolumn{2}{c}{\textbf{Explanation}} & \textbf{Workers} \\
& \textbf{Notes (User)} & \textbf{CRAVE} & \\
\midrule

LLM & 32.89\% & 67.11\% & N/A \\
AMT & 42.34\% & 57.66\% & 10 \\
\bottomrule
\end{tabular}
\caption{Preference for explanation amongst LLM and AMT worker: Both LLM and AMT workers prefer CRAVE explanations over Community Notes.}
\label{tab:preference_comparison}
\end{table}

\begin{figure*}[htbp]
    \centering
    \includegraphics[width=\textwidth]{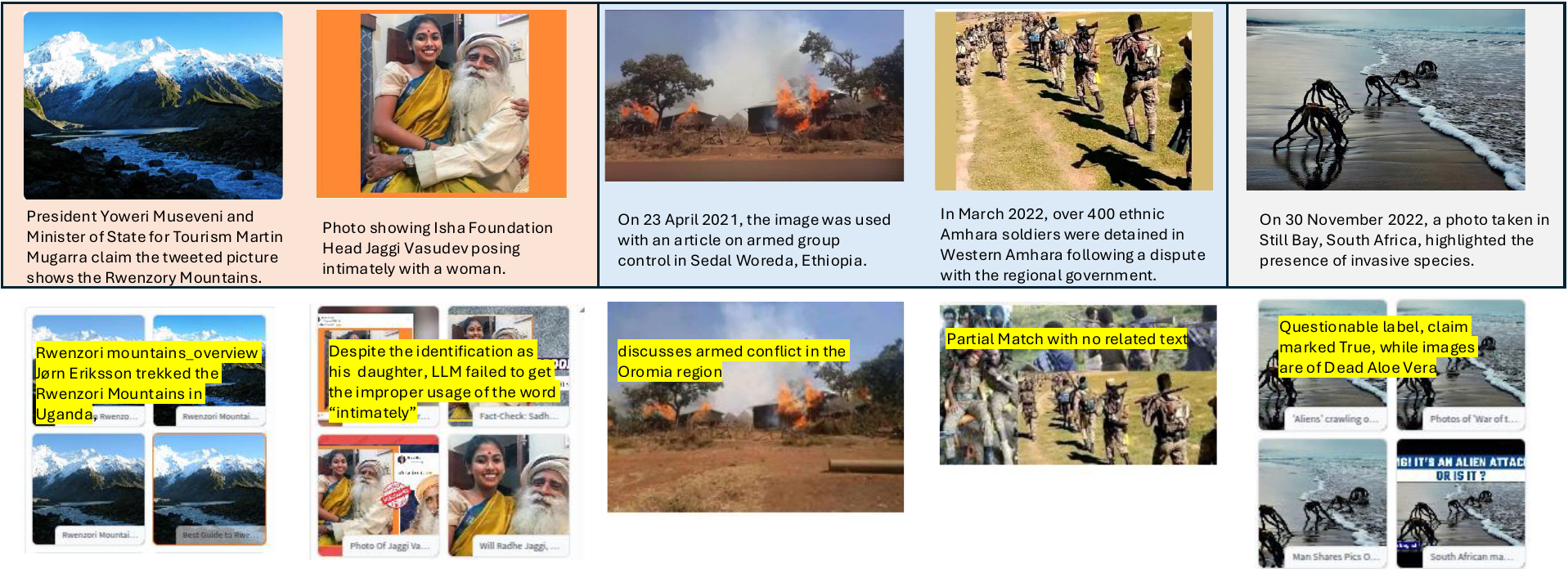}
    \caption{ Misclassified samples are color-coded—orange for those labeled `True', blue for those labeled `Fake'. A sample with a questionable ground-truth label (`True') is shown in grey. Zoom in for details.}
    \label{fig:limitations}
\end{figure*}

\subsection{Critical analysis}

\subsubsection{Analysis 1 - Primary causes of  errors }
\autoref{fig:limitations} presents qualitative examples illustrating key factors behind misclassifications, corresponding to three main types of errors identified in our framework. First, \textbf{clustering issues} occur when an image is widely reused across multiple fabricated contexts, causing the dominant fake narrative to skew evidence grouping and retrieval results (e.g., the leftmost mountain landscape in \autoref{fig:limitations}). Second, \textbf{reasoning errors} arise because the system treats all sources equally without weighing domain credibility, amplifying the influence of low-credibility outlets. While 5W1H-based contextualization helps detect surface-level inconsistencies such as named entities, it often misses subtler linguistic cues needed to disambiguate misinformation, as shown in the second sample of \autoref{fig:limitations}. Third, \textbf{retrieval issues} contribute to failures in verifying true claims, frequently due to scraping limitations that prevent access to relevant textual evidence even when images are successfully retrieved; this is illustrated by the third and fourth examples in \autoref{fig:limitations}. Lastly, some errors stem from questionable dataset labels, as seen in the final example.

Our error analysis reveals three main failure categories, with quantitative results shown in \autoref{tab:error_distribution}.

\begin{enumerate}
    \item \textbf{Retrieval issues (most common):} Errors stem from incomplete retrieval (incom), finding only false evidence (False), or a complete lack of relevant evidence (NO).
    
    \item \textbf{Clustering issues:} True evidence is sometimes grouped with highly similar false narratives (Under-clustering: K$<$). For claims with multiple sub-parts (like in DP dataset), the evidence gets fragmented across clusters (Over-clustering: K$>$).
    
    \item \textbf{Reasoning issues:} The LLM struggles with nuanced actions (Act), missing or vague temporal terms like ``recently'' (Tmp), and non-Western entities (Ent). A high volume of fake narratives (CN) can also overwhelm its final judgment.
\end{enumerate}

\begin{table}[htbp]
\centering
\scriptsize
\setlength{\tabcolsep}{3pt} %
\begin{tabular}{ll|cc|ccccc|ccc}
\toprule
\textbf{Dataset} & \textbf{Label } 
& \multicolumn{2}{c|}{\textbf{Clustering}} 
& \multicolumn{5}{c|}{\textbf{Reasoning}} 
& \multicolumn{3}{c}{\textbf{Retrieval}} \\
                   &            & \textbf{K$<$} & \textbf{K$>$} & \textbf{GT} & \textbf{Act} & \textbf{CN} & \textbf{Ent} & \textbf{Tmp} & \textbf{False} & \textbf{incom} & \textbf{NO} \\
\midrule
\multirow{2}{*}{5PILS\_OOC} & True & 4.4 & 0.0 & 9.2 & 4.8 & 8.3 & 5.7 & 8.7 & 1.7 & 20.1 & 21.4 \\
 & False & 1.3 & 0.0 & 0.9 & 2.2 & 3.9 & 0.0 & 1.7 & 5.2 & 0.4 & 0.0 \\
\midrule
\multirow{2}{*}{DP} & True & 12.1 & 3.8 & 5.7 & 13.4 & 12.1 & 7.6 & 3.8 & 0.0 & 24.2 & 12.1 \\
 & False & 0.0 & 0.0 & 0.6 & 0.0 & 1.3 & 0.0 & 0.0 & 3.2 & 0.0 & 0.0 \\
\bottomrule
\end{tabular}
\caption{Error Type Distribution by Dataset and Label (\%)}\label{tab:error_distribution}
\end{table}

\subsubsection{Analysis 2 - Evidence retrieval failure }

The ``not enough data'' (\textbf{NoD}) predictions occur when our retrieval pipeline fails to find sufficient evidence—often due to niche topics, deepfakes, or paywalled content. In our binary classification task, we conservatively label NoD as ``Misleading.'' These predictions are thus adjudged incorrect when the Ground Truth (GT) is True, and correct when GT is False. In \autoref{tab:nod_analysis} we present our findings. High NoD rates in NewsCLIPpings result from using only the dataset-provided evidence, with no retrieval refinement. While our method does not directly detect deepfake images through image analysis, we address such cases indirectly—either when the content is flagged as deepfake by external sources or when it lacks presence in reputable news outlets. Consequently, in the MMFakeBench dataset, fake content involving deepfake images—which cannot be corroborated through external evidence retrieval—results in the highest NoD  percentage  ($9.71$). 

\begin{table}[htbp]
\centering
\scriptsize
\setlength{\tabcolsep}{3pt} %
\begin{tabular}{lcccc}
\toprule
\textbf{Dataset} & \textbf{\% Total Pred} & \textbf{False Pred which} & \textbf{NoD Pred with} & \textbf{NoD Pred with} \\
\textbf{Name} & \textbf{as False} & \textbf{are NoD} & \textbf{GT True} & \textbf{GT False} \\
\midrule
DP & 59.8 & 5.5 & 2.75 & 2.75 \\
NewsCLIPpings & 61.9 & 10.3 & 4.22 & 6.11 \\
MMFakeBench & 81.7 & 14.8 & 5.06 & 9.71 \\
5PILS\_OOC & 63.1 & 3.3 & 1.80 & 1.47 \\
Verite & 69.9 & 2.1 & 0.99 & 1.11 \\
\bottomrule
\end{tabular}
\caption{Analysis of ``Not Enough Data'' Model Predictions Across Datasets}\label{tab:nod_analysis}
\end{table}

\section{Conclusion}
We introduced CRAVE, a novel framework designed to address multimodal misinformation on social media. CRAVE first organizes retrieved evidence into coherent narrative clusters and then employs an LLM as a judge to assess the veracity of claims based on these narratives. By reasoning over contextualized clusters, CRAVE achieves state-of-the-art performance in zero-shot veracity prediction across multiple datasets, including Verite, MMFakeBench, and 5PilsOOC. Our user study further demonstrates that CRAVE’s explanations are perceived as coherent and comprehensive in many cases, even preferred over community-generated notes. 

\section{Limitations and Future Work}
\label{sec:limitations}

While CRAVE demonstrates a significant step forward in multimodal fact-checking, we acknowledge several limitations that define its current scope and provide clear directions for future research.

\begin{enumerate}
    \item \textbf{Dependencies and Source Trust:} The framework's performance is fundamentally bound by the capabilities and biases of its external dependencies: search engines and Large Language Models. Our reliance on commercial search APIs means the retrieved evidence pool is subject to uncontrollable algorithmic biases (e.g., geographical or political), and we currently \textbf{treat all retrieved sources with equal weight}, making the system susceptible to well-optimized, low-credibility content. This underscores the need for future work in source credibility modeling to be integrated into the evidence-gathering phase.

    \item \textbf{Pragmatic Heuristics:} Our methodology incorporates pragmatic heuristics to ensure broad applicability, most notably the choice of a \textbf{static K=4 for clustering}. Our own analysis (\autoref{fig:clusterscheme}) powerfully illustrates that the optimal cluster count is claim-dependent. We therefore present K=4 not as a universal optimum, but as a \textbf{necessary and empirically grounded baseline} for a dataset-agnostic system. 

    \item \textbf{Computational Cost and Applicability:} The comprehensive nature of our multi-step pipeline introduces a significant \textbf{computational cost and latency} ($\sim$4 minutes serially, $\sim$1 minute in batches). This positions CRAVE not as a tool for real-time content moderation, but rather for deep, offline analyses where rigor is prioritized over speed, such as in \textbf{journalistic investigations, academic research, or platform policy audits}.

    \item \textbf{Scope of Modalities:} Finally, our current work is scoped to text-and-image verification. Future iterations should aim to expand the framework's capabilities to address other critical modalities of misinformation, such as video and audio, to provide a more holistic solution to the information forensics challenge.
\end{enumerate}

\section{Ethics statement}
While our explanation explicitly cites the narratives used with the narrative themselves being annotated with sources used, we emphasize that CRAVE is a decision-support, not an autonomous arbiter of truth, and must be used with an awareness of its potential for error.
\bibliographystyle{IEEEtran}
\bibliography{main}
\end{document}